\documentclass[10pt]{iopart}

\usepackage{iopams} 
\usepackage{graphicx}
\usepackage{bbm}
\usepackage{color}

\usepackage[footnotesize]{caption2}
\setcaptionwidth{.7\textwidth}

\newtheorem{proposition}{Proposition}
\newtheorem{lemma}{Lemma}
\newenvironment{proof}[1][Proof]
    {\begin{trivlist}\item[\hskip \labelsep \textit{#1.}]}
    {$\square$ \end{trivlist}}

\begin{document}

\title{A new approach to quantum backflow}
\author{Markus Penz\dag, Gebhard Gr\"ubl\dag, Sabine Kreidl\dag\ and Peter Wagner\ddag}
\address{\dag\ Department of Theoretical Physics\\University
of Innsbruck\\Technikerstr. 25, A-6020 Innsbruck, Austria}
\address{\ddag\ Department of
Engineering Mathematics, Geometry and Computer Science\\University
of Innsbruck\\Technikerstr. 13, A-6020 Innsbruck, Austria}

\ead{markus.penz@uibk.ac.at}

\begin{abstract}
We derive some rigorous results concerning the backflow operator
introduced by Bracken and Melloy. We show that it is linear
bounded, self adjoint, and not compact. Thus the question is
underlined whether the backflow constant is an eigenvalue of the
backflow operator. From the position representation of the
backflow operator we obtain a more efficient method to determine
the backflow constant. Finally, detailed position probability flow
properties of a numerical approximation to the (perhaps improper)
wave function of maximal backflow are displayed.
\end{abstract}



\section{Introduction and summary}

Let a 1-dimensional free solution of the Schr\"{o}dinger equation
contain positive momenta only, and let $P_x(t)$ be this wave
function's probability (at time $t$) to detect the particle at any
position $>x$. Then $P_x(t)$ starts out from $0$ at time
$t=-\infty$ and tends towards $1$ for $t\rightarrow\infty$.
Because of $\dot{P_x}(t)=j(t,x)$, the (position probability)
current $j(t,x)$ is naively expected to be nonnegative for every
$(t,x)$. Yet there exist positive momentum wave functions such
that the current at, e.g., $x=0$ is negative at certain
intermediate times. In this case the half space probability as a
function of time, i.e., $P_0:\mathbb{R}\rightarrow\left[
0,1\right] $ is not monotonically
increasing.\\

This so called quantum backflow effect seems to have been mentioned
first by Allcock in his work on the time of arrival in quantum
physics \cite{allcock}, while Bracken and Melloy
\cite{bracken+melloy} have given the first detailed account of the
phenomenon in 1994. Allcock presented the backflow effect in order
to disprove the hypothesis that the current at $x=0$ yields the
probability density of arrival times for a free positive momentum
wave packet at $x=0$. Recently it has been shown that the backflow
effect indicates discrepancies among two other proposals of arrival
time densities.\cite{ruggenthaler} More specifically it has been
shown in reference \cite{ruggenthaler} that none of the arrival time
densities, which obey Kijowski's axioms \cite{kijowski}, coincides
with the one of Bohmian mechanics \cite{kreidl}. Furthermore their
average arrival times differ if and only if the wave function in
question leads to backflow, in which latter case the average Bohmian
arrival time precedes that of Kijowski's distributions.
\\

Bracken and Melloy \cite{bracken+melloy} posed the question
whether the backflow of probability is restricted by a stronger
bound than the obvious one given by $1$. Though the existence of
such a stronger bound was not to be expected, they attempted to
numerically compute the smallest upper bound $\lambda$ for the
decrease of $P$. By converting this backflow constant $\lambda$
into the supremum of the spectrum of an integral operator $K$ in
momentum space, surprisingly enough, Bracken and Melloy
approximately found its value to be $0.04$. Meanwhile the
precision of the value of $\lambda$ has been improved by Eveson,
Fewster, and Verch
\cite{eveson} to $0.038452$.\\

In the present work we describe a new approximation method to
determine $\lambda,$ which provides independent confirmation of
the results of \cite{eveson}. Such confirmation is in need since a
rigorous proof for the conjecture $\lambda<1$ is still missing.
The basic idea is to use a decomposition of the integral operator
$K$ into a sum of Fourier transformed multiplication operators. In
this way the method of fast Fourier transform becomes applicable
and $\lambda$ can be approximated with less computational effort.
We obtain an improved value for $\lambda$ of $0.0384517$. As a
byproduct of our numerical computations we approximate the
(perhaps improper) wave function of maximal backflow and we exhibit
some of its more detailed position probability flow properties.\\

The primary goal of this work, however, is to provide some exact
results concerning the integral operator $K$ of Bracken and
Melloy. From a unitary equivalence it will become obvious that $K$
is linear bounded and self adjoint. Then we prove that $K$ is not
compact by showing that $-1$ belongs to the spectrum of $K$ yet it
is not an eigenvalue. We have not been able to conclusively answer
the question whether $\lambda$ is an eigenvalue of $K$ in the
strict mathematical sense. However we shall provide numerical
plausibility that this is indeed the case. A more extensive
discussion of some of our results concerning the backflow
phenomenon is given in reference \cite{penz}.

\section{The backflow constant}

The free Schr\"{o}dinger evolution $U_{t}:L^{2}\left(
\mathbb{R}\right) \rightarrow L^{2}\left(  \mathbb{R}\right)  $ from
time $0$ to time $t\in\mathbb{R}$ is given in the momentum
representation by
\[
\left(  U_{t}\phi\right)  \left(  k\right)  =\phi_{t}\left( k\right)
:=\exp\left(  -\rmi k^2 t\right)  \phi(k).
\]
Here $t$ denotes the rescaled time variable $\hbar
t_{\mbox{\scriptsize phys}} / (2 m)$. Let $\psi_{t}$ denote the
inverse $L^{2}$-Fourier transform of $\phi_{t}$,
i.e.,%
\[
\psi_{t}(x):=\left(  \mathcal{F}^{\ast}\phi_{t}\right)  (x)=\frac{1}
{\sqrt{2\pi}}\int_{-\infty}^{\infty}\exp\left(\rmi kx\right)
\phi_{t}(k) \rmd k.
\]
Let a particle have the momentum space wave function $\phi$ at
time $0$. If $\left\Vert \phi\right\Vert =1$, the probability that
a position measurement at
time $t$ yields a position $x>0$ reads%
\[
P(\phi_{t}):=\int_{0}^{\infty}\left\vert \psi_{t}(x)\right\vert ^{2}
\rmd x=\left\langle
\phi_{t},\mathcal{F}\Pi\mathcal{F}^{\ast}\phi_{t}\right\rangle .
\]
Here $\Pi:L^{2}\left(  \mathbb{R}\right)  \rightarrow L^{2}\left(
\mathbb{R}\right)  $ denotes the orthogonal projection with%
\[
\left(  \Pi f\right)  \left(  x\right)  =\left\{
\begin{array}[c]{cc}
f(x) & \mbox{for }x>0\\
0 & \mbox{for }x<0
\end{array}
\right.  .
\]

If a unit vector $\phi\in L^{2}\left(  \mathbb{R}\right)  $ has its
support contained in $\mathbb{R}_{\geq0}$, i.e., if $\Pi\phi=\phi$,
the probability
$P(\phi_{t})$, according to Dollard's lemma \cite{dollard}, obeys $P(\phi_{t}%
)\rightarrow0$ for $t\rightarrow-\infty$ and
$P(\phi_{t})\rightarrow1$ for $t\rightarrow\infty$. However, the
mapping $t\mapsto P(\phi_{t})$ need not monotonically increase
from $0$ to $1$. Rather it may decrease during several
intermediate time intervals.\cite{bracken+melloy} Thus there exist
momentum space wave functions $\phi\in\mathcal{H}_{+}:=\Pi\left(
L^{2}\left( \mathbb{R}\right) \right)  $ such that
$P(\phi_{s})>P(\phi_{t})$ holds for some $s<t$. For such $\phi$ holds%
\[
\lambda\left(  \phi\right)  :=\sup\left\{
P(\phi_{s})-P(\phi_{t})\mid s,t\in\mathbb{R}\mbox{ with }s<t\right\}
>0.
\]
Unit vectors $\phi\in\mathcal{H}_{+}$ without backflow yield
$\lambda\left( \phi\right)  =0$. We define the backflow constant by
\[
\lambda:=\sup\left\{  \lambda\left(  \phi\right)  \mid\phi\in\mathcal{H}%
_{+}\mbox{ with }\left\Vert \phi\right\Vert =1\right\}  .
\]

Introducing the orthogonal projection
$\widetilde{\Pi}_{t}:=U_{t}^{\ast}\mathcal{F}\Pi\mathcal{F}^{\ast}U_{t}$
we obtain for any unit vector $\phi\in L^{2}\left(
\mathbb{R}\right)$
\[
P(\phi_{s})-P(\phi_{t})=\left\langle \phi,\left(  \widetilde{\Pi}
_{s}-\widetilde{\Pi}_{t}\right)  \phi\right\rangle .
\]
Because of
\[
\widetilde{\Pi}_{s}-\widetilde{\Pi}_{t}=U_{\frac{t+s}{2}}^{\ast}\left(
\widetilde{\Pi}_{\frac{s-t}{2}}-\widetilde{\Pi}_{\frac{t-s}{2}}\right)
U_{\frac{t+s}{2}}
\]
it follows that
\[
\lambda=\sup \left\{ \left\langle \phi,U_{\tau}^{\ast}\left(
\widetilde{\Pi }_{-T}-\widetilde{\Pi}_{T}\right)
U_{\tau}\phi\right\rangle \mid \phi \in\mathcal{H}_{+},\left\Vert
\phi\right\Vert =1,\tau\in\mathbb{R},T\in\mathbb{R}_{>0}\right\}  .
\]
Since the unitary $U_{\tau}$ stabilizes $\mathcal{H}_{+}$ we infer
\[
\lambda=\sup \bigcup_{T > 0} \sigma\left( \Pi B_{T}\Pi\right),
\]
where $B_{T}$ denotes the backflow operator
\begin{equation}
B_{T}:=\widetilde{\Pi}_{-T}-\widetilde{\Pi}_{T},\label{BFOP}
\end{equation}
and $\sigma\left(A\right)$ denotes the spectrum of a linear operator
$A$. This follows from theorem 2, section 8, chapter XI of
\cite{yosida}. Observe the bounds $-id \leq B_{T}\leq id$.\\

Let the one parameter family of unitary dilation operators $V_{\mu}
:L^{2}\left(  \mathbb{R}\right)  \rightarrow L^{2}\left(
\mathbb{R}\right)  $ with $\mu\in\mathbb{R}_{>0}$ be given by
$\left(  V_{\mu}\phi\right)  \left(  k\right) = \sqrt{\mu}\phi\left(
\mu k\right)$. The operators $V_{\mu}$ commute both with $\Pi$ and
with $\mathcal{F} \Pi\mathcal{F}^{\ast}$ and a brief computation
shows that
\[
V_{\mu}U_{t}V_{\mu}^{\ast}=U_{\mu^{2}t}.
\]
From this it follows that%
\[
V_{\mu}\Pi B_{T}\Pi V_{\mu}^{\ast}=\Pi B_{\mu^{2}T}\Pi.
\]
Since the spectrum of an operator is invariant under a unitary
transformation we have the following result, on which our
numerical computation will be based.

\begin{proposition}
\label{BFConst}For any fixed real $T>0$ holds
$\lambda=\sup\sigma\left(\Pi B_{T}\Pi\right).$
\end{proposition}

In view of this result we choose $T=1$ in what follows. The
corresponding operators $U_{T=1}$ and $B_{T=1}$ will be
abbreviated by $U$ and $B$.

\section{Equivalence with the treatment of Bracken and Melloy}

Now we will prove that our definition of $\lambda$ indeed is
equivalent to the one of Bracken and Melloy \cite{bracken+melloy}.
These authors heuristically introduce $\lambda$ via time integrals
of currents at point $x=0$ over arbitrary finite intervals. From
this they motivate their final definition of $\lambda$ as the
supremum of the spectrum of the integral operator
\begin{eqnarray*}
K &:& L^{2}\left(  \mathbb{R}_{>0}\right)  \rightarrow L^{2}\left(
\mathbb{R}
_{>0}\right)  \\
&& \mbox{with}\:\left(  Kf\right)  \left(  k\right) =-\frac
{1}{\pi}\int_{0}^{\infty}\frac{\sin\left( k^{2}-q^{2}\right)
}{k-q}f\left( q\right) \rmd q.
\end{eqnarray*}

Let $\eta:\mathcal{H}_{+}\rightarrow L^{2}\left(
\mathbb{R}_{>0}\right)  $ denote the unitary operator with $\left(
\eta\phi\right)  \left( k\right)  =\phi(k)$ for all $k>0$ and for
all $\phi$ in $\mathcal{H}_{+}$.

\begin{proposition}
For all $\phi\in\mathcal{H}_{+}$ there holds $K\eta\phi=\eta\Pi
B\Pi\phi$, i.e., the restriction of $\Pi B\Pi$ to
$\mathcal{H}_{+}$ and $K$ are unitary equivalent.
\end{proposition}

\begin{proof}
Since $\Pi B\Pi$ is bounded it is sufficient to show $\eta\Pi B
\Pi\phi=K\eta\phi$ for all $\phi$ from a dense subspace $\mathcal{D}
\subset\mathcal{H}_{+}$. We shall choose $\mathcal{D}=\mathcal{S}
_{+}(\mathbb{R})$, the space of all $\mathcal{C}^{\infty}$ functions
from $\mathbb{R}$ to $\mathbb{C}$ with fast decrease and with their
support contained in $\mathbb{R}_{>0}$.\\

As a prerequisite we first demonstrate a relation between the
orthogonal
projection $\mathcal{F}\Pi\mathcal{F}^{\ast}$ and the Hilbert transformation%
\[
H:L^{2}\left(  \mathbb{R}\right)  \rightarrow L^{2}\left(
\mathbb{R}\right) ,\quad\left(  Hf\right)  \left(  k\right)
=\frac{1}{\pi} \, \mathcal{P} \hspace{-0.5em}
\int_{-\infty}^{\infty}\frac{f\left( q\right)  }{k-q} \rmd q.
\]
Here $\mathcal{P}$ indicates that the improper integral is meant as
the principal value. For $f\in\mathcal{S}(\mathbb{R})$ we obtain by
means of Lebesgue's dominated convergence theorem and by means of
Sochozki's formula \cite{wladimirow}
\begin{eqnarray*}
\left( \mathcal{F}\Pi\mathcal{F}^{\ast}f\right)  \left(  k\right) &
=\frac{1}{2\pi}\int_{0}^{\infty} \rme^{-\rmi kx}\left(
\int_{-\infty}^{\infty}
\rme^{\rmi xq}f\left(  q\right) \rmd q\right) \rmd x\\
&
=\frac{1}{2\pi}\lim_{\varepsilon\searrow0}\int_{-\infty}^{\infty}f\left(
q\right)  \left(  \int_{0}^{\infty} \rme^{\rmi\left(  q-k\right)
x-\varepsilon
x} \rmd x\right) \rmd q\\
& =-\frac{1}{2\pi \rmi}
\lim_{\varepsilon\searrow0}\int_{-\infty}^{\infty} \frac{f\left(
q\right)}{q-k+\rmi\varepsilon} \rmd q\\
& =-\frac{1}{2 \pi \rmi}\left\{ \mathcal{P} \hspace{-0.5em}
\int_{-\infty}^{\infty}\frac{f\left(
q\right)  }{q-k}\rmd q-\rmi\pi f\left(  k\right)  \right\} \\
& =\frac{1}{2 \rmi}\left(  Hf\right)  \left(  k\right)
+\frac{f\left( k\right) }{2}.
\end{eqnarray*}
By continuity we infer
\begin{equation}
\mathcal{F}\Pi\mathcal{F}^{\ast}=\frac{1}{2}\left(  -\rmi H+id\right)  .\label{HS}%
\end{equation}
From equation (\ref{HS}) it is easy to show that the Hilbert
transformation is unitary and that $\sigma\left(  H\right)
=\left\{\rmi, -\rmi \right\}$.\\

From the equations (\ref{BFOP}) and (\ref{HS}) follows
\begin{equation}
B=\frac{1}{2\rmi}\left(  UHU^{\ast}-U^{\ast}HU\right).\label{BFHT}
\end{equation}
From this we obtain for $k>0$ and $\phi\in\mathcal{D}$
\begin{eqnarray*}
\left(  \Pi B\Pi\phi\right)  \left(  k\right) & =\frac{\rme^{-\rmi
k^2}} {2 \rmi}\left( HU^{\ast}\phi\right) \left(
k\right)-\frac{\rme^{\rmi k^2}}
{2 \rmi}\left( HU\phi\right) \left(  k\right) \\
& =-\frac{1}{\pi}\int_0^\infty \frac{\sin\left( k^2-q^2\right) }
{k-q}\phi\left(q\right) \rmd q=\left( K\phi\right) \left( k\right) .
\end{eqnarray*}
Clearly for $k<0$ holds $\left(  \Pi B\Pi\phi\right)  \left(
k\right) =0$. By continuity we have $\Pi
B\Pi\phi=\eta^{-1}K\eta\phi$ for all $\phi\in\mathcal{H}_{+}$.
Thus the restriction of $\Pi B\Pi$ to $\mathcal{H}_{+}$ is unitary
equivalent to $K$.\\
\end{proof}

Therefore the defining relation of \cite{bracken+melloy}, $\lambda
=\sup\sigma\left(K\right),$ indeed holds.

\section{Noncompactness}

\begin{proposition}
The backflow operator $\Pi B \Pi$ is not compact.
\end{proposition}

\begin{proof}
For $\Pi B \Pi$ holds $-id\leq\Pi B \Pi\leq id $.\ Therefore
$\sigma\left(  \Pi B \Pi\right) \subset\left[ -1,1\right] $. For
every unit vector $\phi\in\mathcal{H}_{+}$, according to Dollard's
lemma \cite{dollard} holds
\[
\lim_{T\rightarrow\infty}\left\langle \phi,\Pi
B_{T}\Pi\phi\right\rangle =-1.
\]
Since the spectrum of $\Pi B_{T}\Pi$ does not vary with $T$ it
follows that $-1\in\sigma\left(  \Pi B \Pi\right)$. If $\Pi B \Pi$
were compact, then $-1$ were an eigenvalue of $\Pi B \Pi$. Let
$\phi\in L^{2}\left(  \mathbb{R}\right)  $ with $\left\Vert \phi
\right\Vert =1$ denote an eigenvector of $\Pi B \Pi$ with eigenvalue
$-1$, i.e., $\Pi B \Pi\phi=-\phi$ holds. Since $\Pi B \Pi\phi\in
\mathcal{H}_{+}$ it holds that $\phi\in\mathcal{H}_{+}$. Then it
follows from the triangle inequality, from the unitarity of the
Hilbert transformation $H$, and from equation (\ref{BFHT}) that
\begin{eqnarray*}
1  & =\left\Vert \phi\right\Vert =\left\Vert \Pi B
\Pi\phi\right\Vert
=\left\Vert \Pi B \phi\right\Vert \\
& \leq\left\Vert B \phi\right\Vert =\frac{1}{2}\left\Vert \left(
UHU^{\ast}-U^{\ast}HU\right)  \phi\right\Vert \\
& \leq\frac{1}{2}\left(  \left\Vert UHU^{\ast}\phi\right\Vert
+\left\Vert U^{\ast}HU\phi\right\Vert \right)  =\frac{1}{2}\left(
\left\Vert HU^{\ast }\phi\right\Vert +\left\Vert HU\phi\right\Vert
\right)  =1
\end{eqnarray*}
Thus the triangle inequality becomes an equality and we have
\[
\left\Vert \left(  UHU^{\ast}-U^{\ast}HU\right)  \phi\right\Vert
=\left\Vert UHU^{\ast}\phi\right\Vert +\left\Vert
U^{\ast}HU\phi\right\Vert
\]
from which it follows that there exists some $\alpha\in\mathbb{C}$
such that
\[
UHU^{\ast}\phi=\alpha U^{\ast}HU\phi.
\]
Since $UHU^{\ast}$ and $U^{\ast}HU$ are unitary it follows that
$\left\vert \alpha\right\vert =1$. From the above sequence of
inequalities it also follows that
\[
\left\Vert \Pi B \phi\right\Vert =\left\Vert B \phi\right\Vert .
\]
This is equivalent to $\Pi B \phi=B \phi$. Thus the eigenvector
condition $\Pi B \Pi\phi=-\phi$ implies $B \phi=-\phi$, from which
by means of equation (\ref{BFHT}) it follows that
\[
\frac{1}{2\rmi}\left(  \alpha-1\right)  HU\phi=-U\phi.
\]
Thus $U\phi=:\Phi\in\mathcal{H}_{+}$ is an eigenvector of $H$. Since
$\sigma\left(H\right)=\left\{\rmi,-\rmi\right\}  $ it follows that
$\alpha-1\in\left\{2,-2\right\}$. Because of $\left\vert
\alpha\right\vert =1$ this implies $\alpha=-1$ and $H\Phi=\rmi\Phi$.
Thus it follows that
\[
\mathcal{F}\Pi\mathcal{F}^{\ast}\Phi=\frac{1}{2}\left( -\rmi
H+id\right) \Phi=\Phi.
\]
Thus we have $ \Pi \mathcal{F}^{\ast}\Phi=\mathcal{F}^{\ast}\Phi$
for some nonzero $\Phi \in\mathcal{H}_{+}$. Now the following
lemma implies the contradiction $\Phi=0$. Thus $-1\ $is not an
eigenvalue of the backflow operator. Since every nonzero spectral
value of a compact operator is an eigenvalue, the backflow
operator necessarily is noncompact.
\end{proof}

\begin{lemma}
Let $\Phi\in L^{2}\left(  \mathbb{R}\right)  $ with $\Pi\Phi=\Phi$
and $\Pi\mathcal{F}^{\ast}\Phi=\mathcal{F}^{\ast}\Phi$. Then
$\Phi=0$ holds.
\end{lemma}

\begin{proof}
Any function from $L^{2}\left(  \mathbb{R}\right)  $ is locally
integrable. Therefore the inverse Fourier transform of
$\Phi\in\mathcal{H}_{+}$ is the distributional boundary value of the
holomorphic function $\widetilde{\Phi}$
on the complex upper half plane defined by%
\[
\widetilde{\Phi}:\left\{  z\in \mathbb{C}\mid\Im z>0\right\}  ,\quad
a+\rmi b\mapsto \frac{1}{\sqrt{2\pi}}\int_{0}^{\infty}\rme^{\rmi ak}
\rme^{-bk}\Phi(k) \rmd k.
\]
If the boundary value obeys
$\Pi\mathcal{F}^{\ast}\Phi=\mathcal{F}^{\ast}\Phi$, then the
distribution $\mathcal{F}^{\ast}\Phi$ is zero on $\mathbb{R}_{<0}$.
From the generalized uniqueness theorem, see theorem B.10 on p. 100
of \cite{Bogolubov}, it follows that $\widetilde{\Phi}=0$. Thus also
the boundary value $\mathcal{F}^{\ast}\Phi$ of $\widetilde{\Phi}$
vanishes. Since $\mathcal{F}^{\ast}$ is unitary we also have $\Phi =
0$.
\end{proof}

\section{Numerical Computation of the backflow constant $\lambda$}

In \cite{bracken+melloy, eveson} the integral operator $K$ is
approximated by a finite square matrix, whose largest eigenvalue
is taken as an approximation of $\lambda$. If, however, we apply
the power-method to the expression for $\lambda$, which is given
in proposition \ref{BFConst}, we immediately approximate the
largest eigenvalue without having to compute any matrix. One only
needs to apply multiplication operators and fast Fourier
transformations to an arbitrary initial vector.
The power method works as follows. \cite{golub}\\

Let the matrix $A\in\mathbb{C}^{N\times N}$ be symmetric. Let $a$
be the eigenvalue of $A$ with the largest absolute value. Let
$v_{o}\in\mathbb{C}^N$ be a nonzero vector with nonzero component
within the eigenspace of $A$ corresponding to $a$. Then the
sequence $\left(  v_{n}\right)  _{n\in
\mathbb{N}_{0}}$ is recursively defined by%
\[
v_{n+1}=\frac{1}{\left\Vert v_{n}\right\Vert }Av_{n}.
\]
Then holds
\[
a=\lim_{n\rightarrow\infty}v_{n+1}^{\dagger}\cdot\frac{v_{n}}{\left\Vert
v_{n}\right\Vert }.
\]

Since $\sigma\left(  \Pi B \Pi\right)  \subset\left[
-1,\lambda\right]  $ we apply the power method to the nonnegative,
discretized operator $\Pi B \Pi+id$. Its
largest eigenvalue then approximates $\lambda+1$ while $v_n$ tends
towards the corresponding eigenvector.\\

\begin{figure}[ht!]
\begin{center}
\includegraphics[scale=0.5]{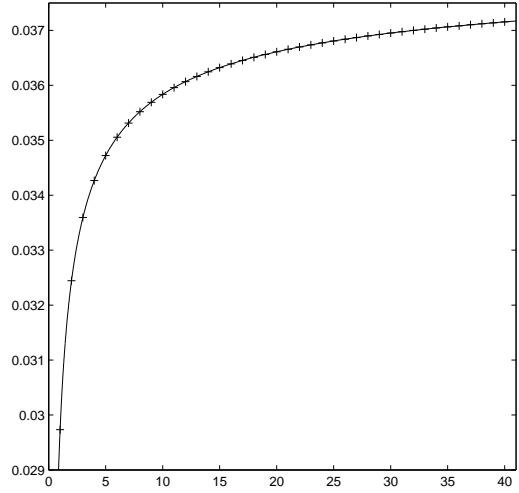}
\caption{$\lambda$ plotted against $h$ and fit $\lambda_\infty +
b/\sqrt{h}$.} \label{fig-meth2-ergeb-1}
\end{center}
\end{figure}

The analysis was started with $N_0 = 10^4$ grid-points covering
the interval $[0, q_0]$, where $q_0$ is set to 50. Now the
power-method was applied with 1000 iterations to a constant
starting vector. Then we repeated the computation for to $N = N_0
h$ grid-points and a larger momentum interval $[0, q]$ with $q =
q_0 \sqrt{h}$ for $h = 1, 2,\ldots40$. In this way the covered
interval grows while the absolute step size $q/N$ gets smaller.
The results $\lambda_h$ for different factors of accuracy $h$ then
were used to extrapolate to $h \rightarrow \infty$ leading to an
approximation $\lambda_\infty$ for the backflow-constant. The
results of this computation can be seen in figure
\ref{fig-meth2-ergeb-1}.\\

In order to check for the possibility that the constant starting
vector $v_{o}$ has vanishing component within the eigenspace of
the dominating eigenvalue various other starting vectors have been
chosen as well. After only few iterations they all led to the same
results. Since it seems extremely unlikely that all chosen
starting vectors have vanishing components within the eigenspace
of the dominating eigenvalue,
our algorithm is likely to approximate the largest spectral value of $\Pi B \Pi+id$.\\

\begin{figure}[ht!]
\begin{center}
\includegraphics[scale=0.5]{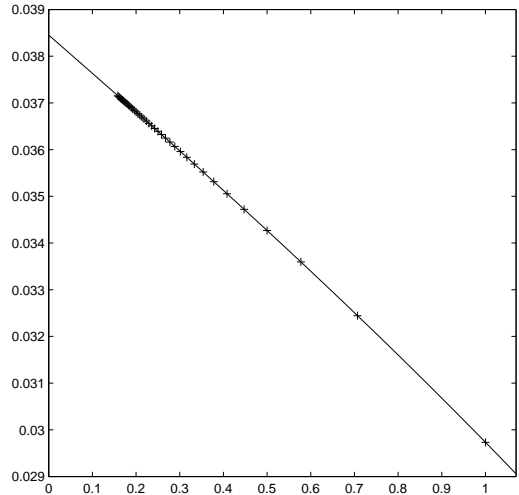}
\caption{$\lambda$ plotted against $1/\sqrt{h}$ and polynomial fit
of third order.} \label{fig-meth2-ergeb-2}
\end{center}
\end{figure}

An even better result for $\lambda_\infty$ is achieved by fitting
the graph of figure \ref{fig-meth2-ergeb-2} with a polynomial of
third order. The extrapolated value for the backflow-constant can
then be read off from the intersection of the y-axis with the
graph. (This corresponds to the point $1/\sqrt{h} = 0$ on the
x-axis and $h \rightarrow \infty$ respectively.) This yields
\[
\lambda_\infty = 0.0384517
\]
which agrees with the value given in \cite{eveson} by $0.038452$.\\

By means of the power-method we also get an approximation of the
possibly improper eigenvector associated with the backflow
constant. It will be discussed briefly in the next section.

\section{Approximate backflow maximizing vector}

Since the operator $K$ of Bracken and Melloy is real, the
(improper?) backflow maximizing eigenvector may be chosen to be
real valued  in the momentum representation. From this it follows
that the position representation at time $0$ has even real part
and and odd imaginary part. More generally, the time dependent
wave function is invariant under the combined parity
and time  reversal operation.\\

\begin{figure}[h!]
\begin{center}
\includegraphics[width = 0.7\textwidth]{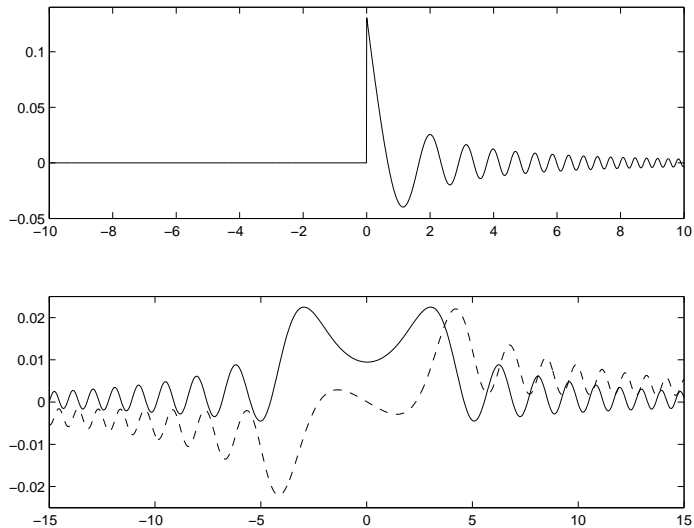}
\caption{$v_n$ in momentum and configuration space (below, real
part \full, imaginary part \broken).} \label{fig-eigenvec}
\end{center}
\end{figure}

We take as an approximate backflow maximizing vector the vector
$v_n$ obtained from the power method, where we choose $N = 10^4$,
$q = 50$ and we make $n=1000$ iterations. The starting vector
$v_0$ is -- as before -- simply the constant function. This leads
-- as one can read off from figure \ref{fig-meth2-ergeb-2} -- to
quite a bad approximation of $\lambda$ by about $0.0297$, but a
further increase of the accuracy leaves the appearance of the
approximate eigenvector $v_n$ as displayed in figure \ref{fig-eigenvec} totally unchanged.\\

\begin{figure}[h!]
\begin{center}
\includegraphics[width = 0.7\textwidth]{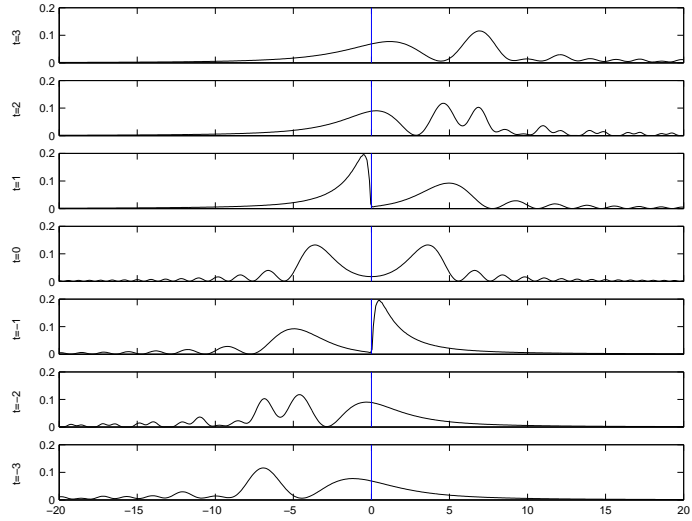}
\caption{Position probability density of $(v_n)_t$ for $-20<x<+20$
at times $t \in \{-3,-2, \dots, +3 \}$.} \label{evol1}
\end{center}
\end{figure}

\begin{figure}[h!]
\begin{center}
\includegraphics[width = 0.8\textwidth]{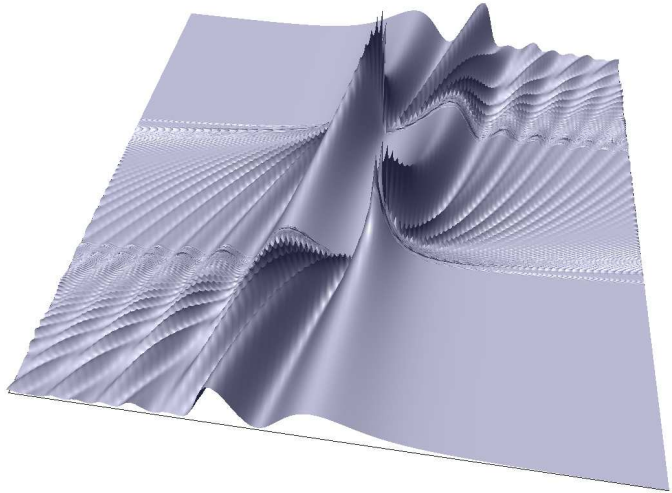}
\caption{Position probability density of $(v_n)_t$ for $-20<x<+20$
and $-3 < t < 3$.} \label{evol2}
\end{center}
\end{figure}

The position probability density of $v_n$ subject to the free time
evolution is displayed in figures \ref{evol1},
\ref{evol2}. These figures by themselves do not provide
unquestionable evidence for the appearance of backflow.\\

\begin{figure}[h!]
\begin{center}
\includegraphics[width = 0.7\textwidth]{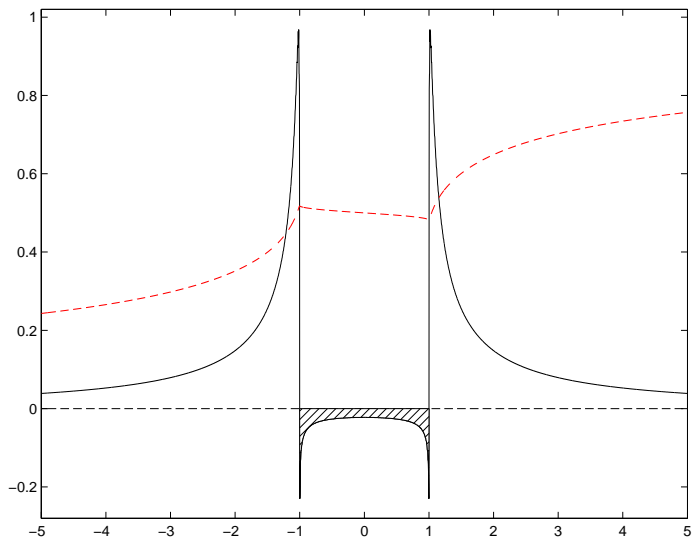}
\caption{$j(t,0)$ (\full) of $v_n$ and the corresponding half
space probability (\textcolor{red}{\broken}) as functions of
time.} \label{fig-eigenvec-j}
\end{center}
\end{figure}

\begin{figure}[h!]
\begin{center}
\includegraphics[width = 0.72\textwidth]{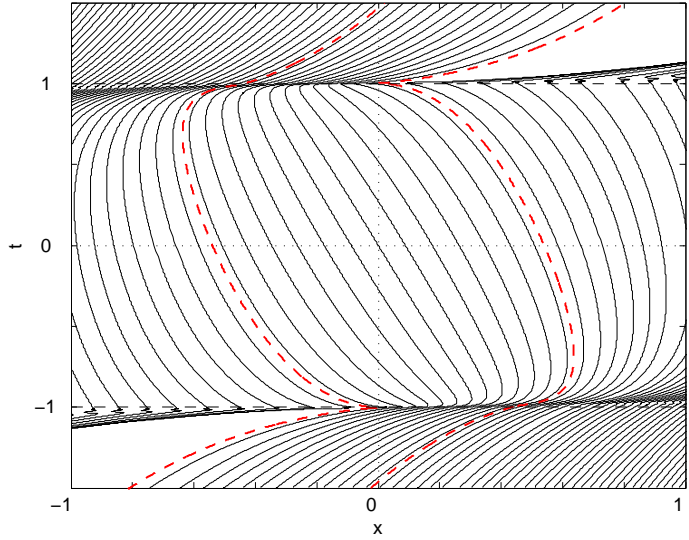}
\caption{Flow lines of the velocity field $j/\rho$ of $v_n$. The
flow lines between \textcolor{red}{\broken} pass through $x=0$ in
the negative direction. Two consecutive flow lines are separated
by a probability of approximately $2.4 \cdot 10^{-3}$.}
\label{flowlines}
\end{center}
\end{figure}

In order to strikingly illustrate the backflow we compute for the
approximate backflow maximizing vector $v_n$ the current $j(t,x)$
at position $x = 0$ as a function of time. The result is shown in
figure \ref{fig-eigenvec-j}, where the backflow-domain $]-1,1[$ is
plainly indentifiable. This interval seems to be the only one in
which $v_n$ leads to a backflow through $x=0$. The area below it,
as required, approximately sums up to the backflow-constant. The
corresponding half space probability as a function of time is also
shown in figure \ref{fig-eigenvec-j}. Further evidence for the
backflow phenomenon of $v_n$ is provided by figure
\ref{flowlines}. This figure shows some integral curves of the
space time vector field $(1,j/\rho)$, the flow lines of the
Bohmian velocity field, within the backflow-domain. All the integral curves
which pass the line $x=0$ at a time $t$ with $-1<t<1$ pass it in the negative direction.\\

\begin{figure}[h!]
\begin{center}
\includegraphics[width = 0.6\textwidth]{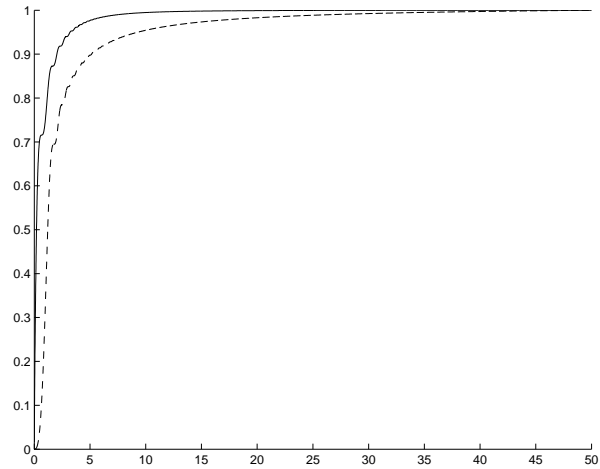}
\caption{Norm-squares of $v_n$ (\full), and $\sin(k^2)/k$
(\broken).} \label{fig-eigenvec-norm-konv}
\end{center}
\end{figure}

The question which remains open is this: Is there really a
backflow-eigenvalue -- in the strict mathematical sense -- to
which $\lambda_\infty$ is an approximation? From the approximate
eigenvector $v_n$ evidence can be found that there is indeed one.
To this end we compute the contribution of the interval $[0,q]$ to
the norm-square of $v_n$ and compare it to $\int_0^q |f(k)|^2 \rmd
k$ with $f(k) = \mathcal{N} \cdot \sin(k^2)/k$ where $\mathcal{N}$
is a normalization constant. Note that $f \in L^2(\mathbb{R})$.
The results are shown in figure \ref{fig-eigenvec-norm-konv}. The
two graphs are very similar and the norm of the $v_n$ seems to
converge even faster than that of $f$. Thus it seems plausible
that $\lambda$ is indeed an eigenvalue of the backflow operator.

\end{document}